# The MSSM compatibility with the limit on electron electric dipole moment

S.S. AbdusSalam[1], S.S. Barzani[1,2], L. Kalhor[1], M. Mohammadidoust[1], S.A. Ojaghi[1]

1. Department of Physics, Shahid Beheshti University, Tehran, Iran
2. Department of Physics, Antwerp University, Antwerp, Belgium

E-mail: abdussalam@sbu.ac.ir

**Abstract:**

The minimal supersymmetric standard model (MSSM) particles can generate loop-level radiative corrections that contribute to the electric dipole moment (EDM) of an electron. The upper bound on the EDM can therefore be used for delineating the MSSM parameters space. We use this setting to describe a direction of particle physics phenomenology research – the global fits of particle physics models beyond the standard model. This is done within the context of the MSSM phenomenology framework with thirty free parameters (MSSM30). Using samples of MSSM30 parameter-space points constrained with the latest bound on the electron EDM, we show that $\mathrm{Arg}(\mu M_2)$ is the most constrained MSSM CP-violating phase. The EDM-compatible parameter regions feature multi-TeV pseudoscalar Higgs bosons and relatively lower *tan* β values compared to previous analyses.

# پدیده‌شناسی ام-اس-اس-ام و حد گشتاور دوقطبی الکتریکی الکترون

شهو عبدالسلام[1]، صفورا صادقی برزانی[1,2]، لیلا کلهر[1]، محمد محمدی دوست[1]، سید امیررضا اجاقی[1]

1. دانشکده فیزیک، دانشگاه شهید بهشتی، تهران، ایران
2. دانشکده فیزیک، دانشگاه آنتورپ، آنتورپ، بلژیک

پست الکترونیکی: Corresponding auhor email: abdussalam@sbu.ac.ir

**چکیده**

ذرات موجود در مدل ابر تقارن کمینه[1] (ام-اس-اس-ام) می‌توانند اصلاحات تابشی در سطح حلقه ایجاد کنند که به گشتاور دوقطبی الکتریکی[2] (ای-دی-ام) ذرات، مانند ذره الکترون، سهم بدهد. بنابراین کران بالایی در ای-دی-ام می‌تواند برای اختصاص دادن فضای پارامترهای ام-اس-اس-ام بکار رود. با این روی کرد به یک جهت تحقیقات پدیدارشناسی فیزیک ذرات بنیادی که برازش جامع مدل‌های ورای مدل استاندارد است معرفی و توصیف می شود. این در چارچوب نظریه پدیدارشناسی ام-اس-اس-ام با سی پارامتر آزاد (ام-اس-اس-ام-۳۰) انجام می‌شود. با استفاده از دسته هایی از داده های نقاط در فضای پرامترهای ام-اس-اس-ام-۳۰ همراه با کران بالایی روی ای-دی-ام الکترون نشان می دهیم که $\mathrm{Arg}(\mu\, M_2)$ محدودترین فاز نقض کننده بار-پاریته می باشد. مناطق پارامتر سازگار با قید در نظر گرفته دارای بوزون‌های هیگز شبه نردهای چند ترا الکترون ولت و مقادیر $\tan \beta$ کمتر نسبت به نتایج تحلیلات قبلی هستند.

**واژه‌هاي کلیدي:** گشتاور دوقطبی الکتریکی، ابرتقارن، عدم تقارن ماده- پاد ماده، آمار بیزی

---

[1] Minimal Supersymmetric Standard Model (MSSM)
[2] Electric Dipole Moment (EDM)

# ۱. مقدمه

از یک دیدگاه، انسان به عنوان جزئی از طبیعت به دنبال این است که بتواند به درک عمیق‌تر و بهتری از طبیعت دست یابد و به نتایج و دست آوردهایی برسد که به نفع بشریت باشد. برای رسیدن به این هدف به مطالعه کارهای نظری و تطبیق آن با کارهای تجربی می‌پردازد.

شکل‌گیری عالم را می‌توان به عنوان یک مثال در نظر گرفت. در نظریه‌ی مه‌بانگ[1] بعد از زمان خیلی کوتاهی از لحظه آفرینش در اثر کاهش دما و انرژی اولیه، ذرات بنیادی شکل گرفتند. همچنان با گذر زمان و سرد شدن کیهان اولیه، به‌ترتیب اتم‌ها، مولکول‌ها، ستارگان و کهکشان‌ها بوجود آمدند. ایده نظری انبساط عالم توسط الکساندر فریدمان در سال ۱۹۲۲ با بررسی تبعات نظریه نسبیت عام پیشنهاد شد. او با تحلیل معادلات کیهانی نشان داد عالم می‌تواند در حال انبساط باشد، اما این تنها یک نظریه بود و لازم بود شواهدی برای اثبات آن یافت شود. در سال ۱۹۲۹ ادوین هابل با کمک بزرگ‌ترین تلسکوپ آن زمان متوجه شد که کهکشان‌ها در حال دور شدن از ما و از یکدیگر هستند، در نتیجه نظریه‌ی فریدمان با مشاهدات هابل تأیید شد.

برای اینکه بتوانیم عالم هستی را به روش تجربی مطالعه کنیم می‌توان از ابزارهایی مانند ابرذره‌بین‌ها[2] و ابردوربین‌ها[3] برای کاویدن عالم در مقیاس‌های بسیار کوچک و بزرگ استفاده می‌کنیم. در مقیاس نظریه‌هایی که در چارچوب ریاضی مدل‌سازی و ساخته می‌شوند، می‌توان به عنوان مثال به مدل استاندارد ذرات بنیادی در چهارچوب نظریه میدان‌های کوانتومی اشاره کرد. پیش‌بینی‌هایی که از یک مدل مدنظر به دست می‌آیند، بایستی با نتایج تجربی مربوطه سازگار باشد.

عالم در مقیاس کوچک، از ذرات بنیادی ساخته شده است که خود به ذرات کوچک‌تر تجزیه نمی‌شوند و ساختار داخلی ندارند. این ذرات بنیادی در جداول مدل استاندارد به عنوان ذرات بنیادی دسته‌بندی شده‌اند. یک ذره بنیادی آشنا الکترون است که در گروه لپتون‌ها قرار دارد. لپتون‌ها و کوارک‌ها (اجزای تشکیل‌دهنده پروتون و نوترون) در دسته‌ی فرمیون‌ها، ذراتی با اسپین نیمه صحیح، قرار می‌گیرند و به عنوان ذرات تشکیل‌دهنده ماده معمول به حساب می‌آیند. دسته دیگر ذرات بنیادی که بوزون‌ها، ذراتی با اسپین صحیح، هستند به عنوان ذرات واسطه می‌باشند که مسئول برهم‌کنش بین ذرات مادی می‌باشند. بوزون‌های $W^{\pm}$ و $Z$، ذرات واسطه در برهم‌کنش‌های هسته‌ای ضعیف، فوتون عامل برهم‌کنش‌های الکترومغناطیسی و گلوئون‌ها مسئول برهم‌کنش‌های هسته‌ای قوی هستند. در نهایت بوزون هیگز در فرآیند جرم‌دار شدن ذرات بنیادی نقش ایفا می‌کند.

مدل استاندارد که خیلی کوتاه به آن اشاره کردیم برای مباحثی مانند انرژی تاریک، ماده تاریک و عدم تقارن ماده و پادماده در کیهان جوابی ندارد و می‌توان نتیجه گرفت که این مدل هنوز کامل نیست. به همین دلیل محققان برای پاسخگویی به سؤالاتی از این قبیل به سراغ مدل‌های ورای استاندارد می‌روند. عالم در حال انبساط است. این درحالی است که بر اثر عاملی که آن را نمی‌شناسیم سرعت این انبساط در حال افزایش است (شتاب‌دار می‌باشد)، این عامل ناشناخته، انرژی تاریک نام دارد. حدود ۷۰ درصد ماده عالم از این انرژی تاریک ناشناخته تشکیل شده است. حدود ۲۶ درصد عالم از ماده تاریک تشکیل شده که به عنوان مولفه‌ای برای تشکیل و ثبات ساختارها در مقیاس بزرگ عالم وجود آن لازم است و ما تنها ۴ درصد عالم پیرامون خود را می‌شناسیم. به نظر می‌رسد در آغاز عالم، تشکیل ذرات از انرژی به دو صورت ماده و پادماده به میزان یکسان بوده است. امروز عدم تقارن ماده و پادماده این سؤال را تداعی می‌کند که چه اتفاقی در دوران آغازین پیدایش عالم افتاده است که جهان امروز تنها از ماده‌ها تشکیل شده است؟ تحقیقات برای شناخت ماده تاریک و منشأ عدم تقارن ماده و پادماده، مثال‌هایی برای گذر از مدل استاندارد به سمت مدل‌های ورای استاندارد هستند که با هدف به دست آوردن جوابی برای این دست از سؤالات مطرح می‌شوند.

با استفاده از مدل‌های مختلف برگرفته از کارهای نظری و داده‌هایی که از آزمایشگاه یا از طریق مشاهدات بدست می‌آیند، می‌توان به شناخت بهتر و دقیق‌تر طبیعت دست یافت. اگر بخواهیم طبیعت را درست و دقیق بررسی کنیم، نیاز است که بتوانیم ارتباط درستی بین کارهای نظری و تجربی داشته باشیم، یعنی بدانیم علم پدیده شناسی ذرات[4] به چه صورت عمل می‌کند. این علم مانند پرنده‌ای است که برای پرواز همزمان به دو بال نظری و تجربی نیاز دارد. لازم است کارهای نظری را با نتایج کارهای تجربی مقایسه کرد تا هم‌خوانی آنها بررسی شود و همین‌طور در رابطه با کارهای تجربی نیاز به یک همکاری هست که به عنوان مثال، گاهی آزمایشی طراحی می‌شود تا یک پیش‌بینی نظری را بیازماید. در این میان کار پدیده‌شناسی ذرات بنیادی مانند پلی ارتباطی بین این دو حوزه است. در شکل(۱)، اگر محور زمان را از پایین به بالا در نظر بگیریم، کارهای نظری که با پدیده شناسی در ارتباط هستند در نهایت کارهایی خواهند شد که خروجی آنها با طبیعت سازگاری بیشتری دارد. در

---

[3] Super-Telescopes
[4] Particle physics phenomenology
[1] Big Bang Theory
[2] Super-Microscopes

رابطه با کارهای تجربی هم وضعیت به همین ترتیب است [۱].

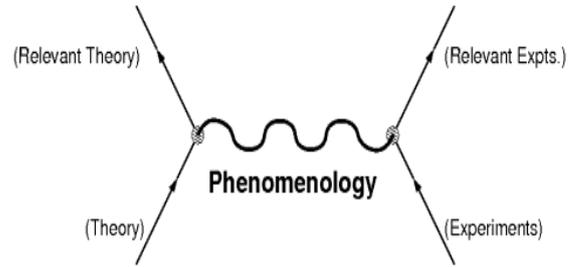

شکل ۱. پدیده‌شناسی به عنوان پلی بین کار تجربی و کار نظری

برای این نوع از تحقیقات پدیده‌شناسی در فیزیک ذرات بنیادی، آگاهی به مهارت‌های مختلف نیاز است. به عنوان مثال آشنایی به مباحث نظریه میدان‌های کوانتومی، مدل‌سازی مربوط به آن، فیزیک محاسباتی، بعضی مباحث آماری، برنامه نویسی رایانه و ابر رایانه و ... لازم است.

در این مقاله روند کار تحقیقاتی پدیده شناسی ذرات بنیادی مرور می‌شود. طوری که برای این منظور، نظریه ابرتقارن و خصوصاً مدل ام-اس-اس-ام در مقایسه با نتیجه تجربی مربوط به جستجوی گشتاور دو قطبی الکتریکی الکترون در نظر گرفته شده است. در بخش بعدی نحوه کاربرد روش آماری بیزین[1] در پدیده شناسی و برازش کلی مدل‌ها معرفی می‌شود. در بخش سوم این مقاله، گشتاور دو قطبی الکتریکی الکترون از دید نظری و تجربی مورد بررسی قرار می‌گیرد و به اهمیت این موضوع در حوزه کیهان‌شناسی، مباحث مربوط به جهان اولیه و ارتباط تنگاتنگ این دو با حوزه‌ی ذرات بنیادی می‌پردازیم. در بخش چهارم، مدل ابرتقارن ام-اس-اس-ام، را معرفی کرده و با استفاده از پدیده‌شناسی بحث حد گشتاور دوقطبی الکتریکی الکترون، فضای پارامتری مدل ام-اس-اس-ام۳۰ را بررسی می‌کنیم. در نهایت به جمع‌بندی و نتیجه‌گیری تحقیقات انجام شده می‌پردازیم.

## ۲. روش آماری بیزین

قضیه‌ی بیزین را می‌توان برای مقایسه پیش‌بینی‌های مدل با مقادیر تجربی تعیین شده مشاهده‌پذیرهای خاص مورد استفاده قرار داد. ابتدا لازم است که پارامترهای مدل، θ، و تابع توزیع اولیه، $p(\theta)$، مسأله را مشخص کنیم. عبارت $p(\theta)$ که به صورت توزیع احتمال پیشین[2] معرفی می‌شود، به این معنا است که درک عمیق‌تری نسبت به پارامترهای مدل قبل

با در نظر گرفتن مشاهدات تجربی، قبل از اینکه مدل با داده‌های تجربی فعلی مقایسه شود به ما ارائه می‌کند. شکل تابع $p(\theta)$ معمولاً از کارهای نظری و یا نتایج حاصل از تحلیل‌های قبلی برازش بیزین مدل بدست می‌آید. سپس، مجموعه‌ای از کمیت‌های قابل مشاهده را برای انجام تجزیه و تحلیل پدیدارشناختی انتخاب می‌کنیم و مقادیر آنها را به عنوان عبارت‌های اندازه‌گیری شده از آزمایش‌ها با مجموعه‌ای، d، معین می‌کنیم. d سری داده‌هایی است که به عنوان داده‌های مشاهده‌پذیر در نظر گرفته می‌شوند؛

$$d = \{x_1, x_2, \ldots, x_n\} \quad (1)$$

این‌ها داده‌هایی هستند که از کارهای تجربی بدست می‌آیند. از منظر نظری، با توجه به مجموعه‌ای از مقادیر خاص برای پارامترهای θ، مدل مورد بررسی باید بتواند مجموعه‌ای از پیش‌بینی‌ها را برای مجموعه انتخاب شده از مشاهده‌پذیر ارائه دهد. اجازه دهید مجموعه‌ی پیش‌بینی‌ها را با $d' = \{x'_1(\theta), x'_2(\theta), \ldots, x'_n(\theta)\}$ نمایش دهیم. در حقیقت عناصر این مجموعه غالباً از طریق روش‌های محاسباتی به دست می‌آیند. با این اوصاف ما دو دسته داده خواهیم داشت؛ نخست، داده‌هایی که از طریق کار تجربی بدست می‌آیند، d، و دیگری، داده‌هایی که از طریق روش محاسباتی بدست می‌آیند، $d'$. در نهایت ما این دو دسته از داده‌ها را مقایسه می‌کنیم تا مشخص گردد که داده‌های بدست آمده از روش محاسباتی به چه میزان با داده‌های تجربی هم‌خوانی و مطابقت دارند. برای یک θ معین و مجموعه اندازه‌گیری‌های d، تابع چگالی احتمال $p(\theta|d)$ نشان می‌دهد که پیش‌بینی‌های مدل $d'$ چقدر به d نزدیک هستند. این جمله را به عنوان تابع درست‌نمایی[3] می‌شناسیم که در حالت ایده‌آل، باید به عنوان بخشی از نتایج آزمایش‌ها منتشر شود. در غیر اینصورت، می‌توان آن را به شکل تقریبی یا از طریق مدل‌سازی آماری ساخت [۲]. هنر و زیبایی در تحلیل‌های بیزین این است که پیچیدگی $p(\theta)$ قبلی با احتمال $p(d|\theta)$ چگالی احتمال پسین[4] $p(\theta|d)$ مانند رابطه‌ی $P(\theta|d) \propto P(\theta|d)P(\theta)$ می‌دهد. شامل جمله ثابت تناسب $P(d) = \int P(\theta|d)P(\theta)d\theta$ است که به این ترتیب قضیه بیزین را به صورت به دست می‌دهد.

$$P(\theta|d) = \frac{P(\theta|d)}{P(d)} P(\theta) \quad (2)$$

بنابراین می‌توان یک مدل را با کار تجربی مقایسه کرد و یا اینکه به مقایسه مدل‌های مختلف با یکدیگر پرداخت. جمله $P(\theta|d)$ تابع احتمال پسین[5] نام دارد که از حاصلضرب تابع درست نمایی و احتمال پیشین بدست می‌آید. به عبارت دیگر، احتمال پارامترهای مدل بعد از اینکه از داده‌های تجربی استفاده شده را نشان می‌دهد. $P(d)$ در مخرج عبارت استفاده

---

[1] Baysian Statistical Theory
[2] Prior Probability Distribution
[3] likelihood

[4] Posterior Probability Distribution
[5] Posterior Probability Distribution

شده تا احتمال پسین دارای خصوصیاتی تابع احتمال باشد، مثلاً مقدار آن بین صفر تا یک قرار بگیرد. پس مخرج کسر وظیفه بهنجارسازی و یا سازگاری محاسبات با مفهوم احتمال را دارد. اگر بخواهیم مدل‌های مختلف را مقایسه کنیم، مخرج عبارت قابل استفاده می‌شود. اگر هدف ما تخمین پارامترهای مدل باشد، جمله $P(d)$ حائز اهمیت نیست و به رابطه زیر می‌رسیم؛

(۳)
$$P(\theta|d) \propto P(\theta|d) \times P(\theta)$$

سمت راست معادله (۳) مربوط به ترکیب مدل نظری و کار تجربی است. سمت چپ نتیجه‌ی استفاده از کار تجربی برای تخمین فضای پارامتری مدل است.

با استفاده از مجموعه تعاریف بالا، اکنون می‌توانیم توضیح دهیم که تناسب جهانی بیزین یک مدل با داده‌ها چیست. این فرآیند تکراری یک کاوش هدایت‌شده برای پارامترهای مدل به شیوه‌ای همگرای آماری است که $P(\theta|d)$ را به عنوان یک نتیجه تولید می‌کند. در قسمت بعدی سراغ یک مثال می‌رویم و با این مثال یکی از کاربردهای روش بیزی در فیزیک ذرات بنیادی را نشان می‌دهیم. برای این مثال هم صحبت از کار نظری می‌کنیم و هم به کار تجربی اشاره می‌کنیم و درنهایت یک جمع‌بندی در رابطه با پدیده‌شناسی این مثال با استفاده از مدل ابرتقارن خواهیم داشت. به عنوان مثال در این زمینه می‌توان به مقاله‌های [۶]، [۱۱ و ۱۲] اشاره کرد که با روش بیزی انجام یافته‌اند. در این مقاله، ما بر چگونگی استفاده از $P(\theta|d)$ برای استنتاج نتیجه‌گیری در مورد مدل ام-اس-اس-ام۳۰ با توجه به نتایج آزمایش‌ها در جستجوی گشتاور دو قطبی الکتریکی الکترون تمرکز می‌کنیم. با استفاده از مجموعه اصطلاحات در این بخش، کران بالای ای-دی-ام الکترون حاصل از آزمایش‌ها را به طور خلاصه در بخش بعدی به عنوان اندازه‌گیری برای کاوش و نقشه‌برداری فضای پارامتری ام-اس-اس-ام۳۰ توضیح می‌دهیم.

## ۳. گشتاور دوقطبی الکتریکی[1] الکترون

تاکنون با توجه به نتایج تجربی، به نظر می‌آید که در طبیعت تقارنی به نام سی-پی-تی وجود دارد. قوانین فیزیک تحت تبدیلات همزمان هم‌یوغ بار[2]($C$)، پاریته[3]($P$)، و وارونگی زمانی[4]($T$) ناوردا هستند. با فرض اینکه این تقارن برقرار باشد گشتاور دوقطبی الکتریکی الکترون را بررسی می‌کنیم. باید توجه داشته باشیم نقض $CP$ یعنی نقض $T$ و برعکس. پس نقض $CP$ در یک فرآیند فیزیکی نشان‌دهنده نقض $T$ نیز می‌باشد و می‌تواند جوابی برای این سؤال باشد که چرا در عالمی که امروزه مشاهده می‌کنیم پادماده وجود ندارد. در

آزمایشگاه‌ها که به دنبال نقض سی-پی در طبیعت هستند ویژگی‌های الکترون در ابعاد بسیار کوچک مطالعه می‌شود.

اگر الکترون را به عنوان یک ذره نقطه‌ای در نظر بگیریم، دارای بار الکتریکی منفی واحد است و اگر آن را به شکل یک کره با یک شعاعی در نظر بگیریم، بار الکتریکی آن می‌تواند به صورت کاملاً یکنواخت در سطح این کره توزیع شده باشد. در هر دو صورت گشتاور دوقطبی الکتریکی صفر خواهد بود. تصحیحات کوانتومی می‌تواند به الکترون شکل کروی کمی قطبیده دهد. اگر تصحیح تابشی[5]، که تقارن سی-پی را نقض می‌کند در نظر بگیریم، در این صورت توزیع ناهمگن از بار در سطح کره بوجود می‌آید که می‌توان این بار اضافی کوچک را با $\delta^+$ و $\delta^-$ نشان داد.

مقادیر $\delta^+$ و $\delta^-$ باعث ایجاد گشتاور دوقطبی الکتریکی می‌شود که آن را با $D$ نشان می‌دهیم. در شکل(۲) گشتاور دو قطبی کره الکترون در حالت اصلی با اسپین الکترون موازی است. اگر روی این سیستم تبدیلات مربوط به پاریته $P$، و وارونگی زمانی $T$ را انجام دهیم، تبدیل $P$ باعث می‌شود که جهت گشتاور دو قطبی عوض شود، پس تقارن $P$ نقض می‌شود. تبدیل $T$ جهت اسپین را عکس می‌کند، و در این حالت تقارن وارونگی زمانی نقض می‌شود. در نهایت با توجه به اینکه تقارن سی-پی-تی باید همزمان برقرار باشد، پس تقارن سی-پی نقض می‌شود، این به این معنی است که مشاهده‌ی ای-دی-ام الکترون است. کارهای آزمایشگاهی که در این حوزه انجام می‌شود به دنبال این هستند که مشخص کنند آیا الکترون ذره‌ای نقطه‌ای یا یک کره با توزیع بار الکتریکی همگن و یا حتی یک کره با توزیع بار الکتریکی ناهمگن است یا خیر؟ [۴، ۱۳]

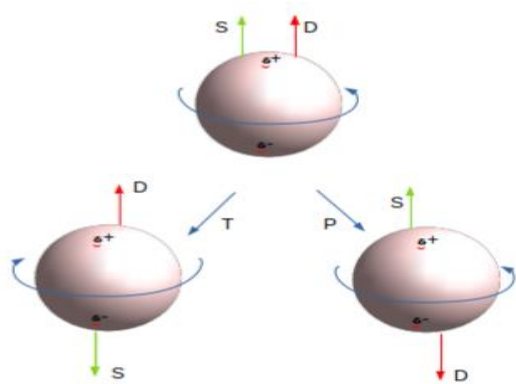

شکل ۲. نمایش گشتاور دوقطبی الکتریکی الکترون و اسپین آن، تحت تبدیل پاریته و وارونگی زمانی

---

اگر الکترون به صورت یک کره با بارالکتریکی ناهمگن در طبیعت مشاهده شود، نشان‌دهنده فیزیکی با نقض سی-پی است. این فیزیک می‌تواند منشاء عدم تقارن ماده و پادماده باشد. به عبارت دیگر گشتاور دوقطبی الکتریکی الکترون با عدم تقارن ماده و پادماده مرتبط است. در نتیجه از طریق مطالعه گشتاور دوقطبی الکتریکی الکترون می‌توان اطلاعاتی مربوط به عالم اولیه به دست آورد که ارتباطی بین بحث فیزیک ذرات بنیادی و کیهان‌شناسی به وجود می‌آید.

از دید نظری، برهمکنش‌های ضعیف در مدل استاندارد ذرات بنیادی مربوط به بوزون $W^\pm$، باعث نقض سی-پی می‌شوند ولی شدت آن برای ایجاد گشتاور دوقطبی الکتریکی الکترون خیلی ضعیف است و همان‌طور که در شکل(۳) نشان داده شده است حداقل تا مرتبه چهار حلقه باید به محاسبه پرداخت. لازم به توضیح است که پیش‌بینی مدل استاندارد در این رابطه چندین مرتبه کمتر از دقت تجربی فعلی است [۱۴ و ۱۵].

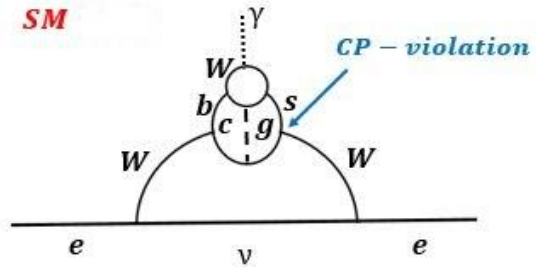

شکل ۳. تصحیح تابشی تا مرتبه ۴ حلقه در مدل استاندارد. با وجود این برهمکنش یک توزیع ناهمگن از بار در سطح الکترون بوجود می‌آید که باعث قطبیده شدن آن می‌شود.

از آنجایی که شدت نقض سی-پی در مدل استاندارد ضعیف است و این شدت برای توضیح عدم تقارن ماده و پادماده کفایت نمی‌کند [۱۶] پس به سراغ مدل‌های فرا استاندارد می‌رویم چرا که در چنین مدل‌هایی شدت نقض سی-پی می‌تواند خیلی بیشتر باشد. به عنوان مثال در شکل(۴)، یک نمودار فاینمن مربوط به مدل ابرتقارن [۳] و [۱۷-۱۹] که تصحیح تابشی تا مرتبه تک حلقه و نقض سی-پی را ایجاد می‌کند، نشان داده شده است. در این شکل دو ذره جدید جفت الکترون($\tilde{e}$)[1] و جفت فوتون($\tilde{\gamma}$)[2] که ذرات مربوط به مدل ابرتقارن هستند، دیده می‌شوند. این مثال مربوط به کار نظری است که در آن الکترون می‌تواند گشتاور دوقطبی الکتریکی داشته باشد.

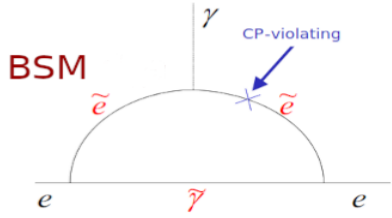

شکل ۴. تصحیح تابشی تا مرتبه تک حلقه در مدل ابرتقارن

تا این قسمت فقط به مباحث نظری اشاره کردیم. در کار تجربی، یک الکترون در داخل یک اتم محبوس می‌شود و مطالعات پیچیده در حوزه لیزر و مولکولی انجام می‌شود. گروه ACME[3] و گروه JIRA[4] که به دنبال یافتن گشتاور دوقطبی الکتریکی الکترون هستند [۴] و [۱۳] یک حد بالا برای این کمیت بدست آورده‌اند:
(۴)

$$\begin{cases} |d_e| < 1.1 \times 10^{-29} \text{e.cm} &, [\text{ACME}] \\ |d_e| < 4.01 \times 10^{-30} \text{e.cm} &, [\text{JIlA}] \end{cases}$$

حال به سراغ روش بیزین می‌رویم تا بتوانیم ارتباطی بین کار نظری و تجربی برقرار کنیم. در این مثال فقط یک داده مشاهده‌پذیر داریم که همان حد بالایی گشتاور دوقطبی الکتریکی است {$d_e$} = d. می‌توان از یک تابع درست‌نمایی ساده مثل تابع پله‌ای برای بررسی استفاده کرد. در شکل(۵) تابع درست‌نمایی، $P(\theta|d)$، نشان داده شده است. سمت راست تابع بر روی عدد یک ثابت شده است و ناحیه‌ای را نشان می‌دهد که مدل و پیش‌بینی نظری با داده تجربی سازگاری داشته باشد. سمت چپ نمودار بر روی مقدار صفر ثابت شده است و حالتی را نشان می‌دهد که مدل با کار تجربی هماهنگی ندارد. با استفاده از دستور بیزی، معادله‌ی (۲)، تابع درست‌نمایی می‌تواند به تابع پسین تبدیل شود. این یک مثال از یک داده مشاهده‌پذیر و یک تابع درست‌نمایی ساده(تابع پله‌ای) است.

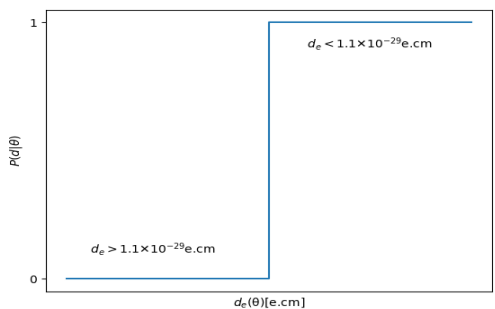

شکل ۵. تابع درست‌نمایی پله‌ای، زمانی که یک مشاهده‌پذیر مثل گشتاور دو قطبی الکتریکی به عنوان داده داریم. به ازای

---

[1] Scalar electron (selectron)
[2] Photino
[3] ACME Collaboration
[4] JIRA Collaboration

پارامترهای مختلف مدل، $d_e$ به دست می‌آید. که اگر مقدار آن از معادله (۴) پیروی کند در این صورت مقدار تابع پله‌ای یک می‌باشد و در غیر این صورت مقدار آن صفر است.

مثال‌هایی وجود دارد که در آن‌ها مشاهده‌پذیرها بیشتر و تابع درست‌نمایی پیچیده‌تر است. می‌توان داده‌های مربوط به جرم هیگز، واپاشی بعضی از هادرون‌ها، نتایج مختلف کارهای تجربی از جمله ال-اچ-سی[1]، نتایج مربوط به ماده تاریک را همگی کنار هم در نظر گرفت و یک مجموعه‌ای از داده مشاهده‌پذیر داشت. محاسبه تابع درست‌نمایی گاهی دشوار است و برخی از گروه‌های تجربی این تابع را محاسبه می‌کنند و در اختیار گروه‌های نظری قرار می‌دهند [۵]. در برخی موارد هم گروه‌های پدیده‌شناسی تابع درست‌نمایی مربوط به کار خودشان را به طور تقریبی تخمین می‌زنند که می‌توان به عنوان مثال به مقاله‌های [۶، ۱۱ و ۱۲] اشاره کرد.

در بخش بعدی، مثال مربوط به حد گشتاور دوقطبی الکتریکی را با استفاده از مدل ابرتقارن به صورت پدیده‌شناسی مطالعه می‌کنیم.

## ۴. پدیده‌شناسی گشتاور دوقطبی الکتریکی در چارچوب مدلی از نظریه ابرتقارن[2]

مدل ابر تقارن یک چارچوب نظری جالب از نظر ریاضی برای ساخت مدل‌های ورای استاندارد فراهم می‌کند. این نظریه براساس بسط تقارن پوانکاره[3] در نظریه میدان کوانتومی توسط یک تقارن فضا-زمان جدید با پیامدهای پدیدارشناختی ساخته شده است[۲۰]. در نظریه ابرتقارن برای هر بوزون در مدل استاندارد ذرات بنیادی یک فرمیون متناظر وجود دارد و بالعکس. همچنین این نظریه می‌تواند برای ساخت مدل‌هایی برای شناخت ماده تاریک مورد استفاده قرار گیرد.

مدل استاندارد ابرتقارن حداقلی[4] (ام-اس-اس-ام) بسطی از مدل استاندارد است و فقط تعداد حداقل ذرات جدید و برهمکنش‌های جدید سازگار با پدیده‌شناسی را در نظر می‌گیرد. برهمکنش این ابرذرات با خودشان و با ذرات مدل استاندارد می‌تواند منشاء نقض سی-پی باشد. بنابراین رابطه (۵) که نتیجه یک کار آزمایشگاهی است، می‌تواند یک قیدی برای پارامترهای ام-اس-اس-ام باشد. یک مدل ام-اس-اس-ام با سی پارامتر آزاد به نام ام-اس-اس-ام۳۰ وجود دارد که پارامترهای این مدل با استفاده از مباحث تقارن، نقض سی-پی و فیزیک طعم تعریف و ساخته شده‌اند [۷-۱۱].

در این بخش، ما به طور خلاصه چارچوب پدیدارشناسی ام-اس-اس-ام۳۰ را توضیح می‌دهیم و سپس به چگونگی بهبود

کران بالایی می‌پردازیم. حد ای-دی-ام برای الکترون [۱۳] پارامترهای مربوط به ام-اس-اس-ام۳۰ را محدودتر می‌کند.

### ۴/۱ مدل حداقلی ابرتقارن با سی پارامتر آزاد، ام-اس-اس-ام۳۰

ام-اس-اس-ام حدوداً دارای صد پارامتر آزاد از شکست ابرتقارن است. رویکردی برای بررسی پارامترها با کاهش تعداد آن‌ها توسط اعمال حداقل نقض طعم[5] (ام-اف-وی) بر روی ام-اس-اس-ام وجود دارد. در رابطه با ساختار ام-اس-اس-ام۳۰، هدف ایجاد معیاری برای حذف کردن اندرکنش‌هایی است که مثلاً مقدار تغییر طعم جریان خنثی در آن‌ها بیشتر است [۱۱]. جمله‌های مربوط به شکست ابرتقارن به دو صورت هستند، جملات جرمی و جملاتی دارای ضرایب یوکاوا. در این مدل مثل مدل استاندارد منشاء نقض طعم ضرایب یوکاوا هستند. این جملات را با ضرایب یوکاوا می‌توان به صورت زیر بسط داد:

$$(M_Q^2)_{ij} = M_Q^2 \left[ \delta_{ij} + b_1(Y_U^\dagger Y_U)_{ij} + b_2(Y_D^\dagger Y_D) \right. \\ \left. + c_1\{(Y_D^\dagger Y_D Y_U^\dagger Y_U)_{ij} + H.C\} + \cdots \right]$$
(۵)

$$(M_U^2)_{ij} = M_U^2 \left[ \delta_{ij} + b_3(Y_U^\dagger Y_U)_{ij} + \cdots \right]$$

$$(M_D^2)_{ij} = M_D^2 [\delta_{ij} + [Y_D(b_6 + b_7 Y_U^\dagger Y_U)Y_D^\dagger]_{ij} + \cdots]$$

$$(M_L^2)_{ij} = M_L^2 \left[ \delta_{ij} + b_{13}(Y_E^\dagger Y_E)_{ij} + \cdots \right]$$

$$(M_E^2)_{ij} = M_E^2 \left[ \delta_{ij} + b_{14}(Y_E^\dagger Y_E)_{ij} + \cdots \right]$$

از طرفی برای پارامترهای دیگر نیز داریم؛

$$(A'_E)_{ij} = a_E \left[ \delta_{ij} + b_{15}(Y_E^\dagger Y_E)_{ij} + \cdots \right]$$
(۶)

$$(A'_U)_{ij} = a_U \left[ \delta_{ij} + b_9(Y_U^\dagger Y_U)_{ij} + b_{10}(Y_D^\dagger Y_D) + \cdots \right]$$

$$(A'_D)_{ij} = a_D \left[ \delta_{ij} + b_{11}(Y_U^\dagger Y_U)_{ij} + b_{12}(Y_D^\dagger Y_D) + \right. \\ \left. c_6(Y_D^\dagger Y_D Y_U^\dagger Y_U)_{ij} + \cdots \right]$$

---

1 Large hadron collider
2 Supersymmetry
3 Poincare Symmetry
4 Minimal Supersymmetric Standard Model
5 Minimal Flavour Violation

در اینجا $M^2_{Q,U,D,E}$ به عنوان ضرایب مربوط به جملات جرمی اسکالرها و $A'_{U,D,E}$ ضرایب اسکالر مربوط به جملات سه تایی است. این جملات قسمتی از شکست ابرتقارن[1] مدل ام-اس-اس-ام را تشکیل می‌دهند.

معادله‌های (۶) و (۷)، به دلیل روابط Cayley-Hamilton برای ماتریس‌های ۳ در ۳ بی‌نهایت نیستند. بنابراین تعداد محدودی از ضرایب $b_k$ و $c_k$ وجود خواهد داشت. به طور کلی برای بسط ماتریس‌ها، مقادیری که ضرایب می‌توانند بگیرند، مرتبه‌های متفاوتی را دربرمی‌گیرند. این جایی است که قدرت فرضیه ام-اف-وی نهفته است چرا که ام-اف-وی بر روی ضرایب قیدی می‌گذارد که باید $b_k$ و $c_k$ از مرتبه یک باشند. حاصل‌ضرب ماتریس‌های یوکاوا که در روابط (۵) و (۶) ظاهر می‌شوند، تابعی از ماتریس‌های پایه که آن‌ها را $X_k$ نامگذاری می‌کنیم:

(۷)
$$X_1 = \delta_{3i}\delta_{3j}, X_2 = \delta_{2i}\delta_{2j}, X_3 = \delta_{3i}\delta_{2j}$$
$$X_4 = \delta_{2i}\delta_{3j}, X_5 = \delta_{3i}V_{3j}, X_6 = \delta_{2i}V_{2j}$$
$$X_7 = \delta_{3i}V_{2j}, X_8 = \delta_{2i}V_{3j}, X_9 = V^*_{3i}\delta_{2j}$$
$$X_{10} = V^*_{2i}\delta_{2j}, X_{11} = V^*_{3i}\delta_{2j}, X_{12} = V^*_{2i}\delta_{3j}$$
$$X_{13} = V^*_{3i}V_{3j}, X_{14} = V^*_{2i}V_{2j}$$
$$X_{15} = V^*_{3i}V_{2j}, X_{16} = V^*_{2i}V_{3j}$$

این‌ها ماتریس‌های پایه از مرتبه یک هستند. در اینجا V ماتریس CKM مدل استاندارد است و $\delta_{ij}$ ماتریس واحد کرونکر می‌باشد. هر یک از عبارت‌های معادله‌های (۵) و (۶) را می‌توان برحسب ماتریس‌های X در معادله‌ی (۷) ضرب در برخی عوامل عددی بیان کرد. با این روش هر یک از جملات $M^2_{Q,U,D,E}$ و $A'_{U,D,E}$ بسته به اینکه با چه ضریبی از عبارت‌های $X_k$ همراه باشند، می‌توانند با مرتبه‌ای از λ طبقه‌بندی شوند. طوری که در اینجا $\lambda = 0.23 = \sin\theta_c$ می‌باشد و $\theta_c$ زاویه کبیبو(Cabibbo) است. هنگامی که دقت محاسبات را برحسب مراتب λ، $O(\lambda^n); n = 1, 2, 3, ...$ مرتب کنیم، در معادله‌های (۵) و (۶) برخی از عبارت‌های موجود را می‌توان به صورت سیستماتیک از بسط بیان شده در مبنای $X_k$ کنار گذاشت و با انتخاب $n = 6$، تعداد پارامترهای ام-اس-اس-ام به ۴۲ کاهش می‌یابد که عبارتند از؛

(۸)
$$\tilde{M}_1 = e^{i\phi_1}M_1$$
$$\tilde{M}_2 = e^{i\phi_2}M_2$$
$$M_3, \tan\beta$$
$$\tilde{\mu} = e^{i\phi_\mu}\mu$$
$$M^2_U = \tilde{a}_2\mathbf{1} + x_2 X_1$$
$$M^2_Q = \tilde{a}_1\mathbf{1} + x_1 X_{13} + y_1 X_1 + y_2 X_5 + y^*_2 X_9$$
$$M^2_D = \tilde{a}_3\mathbf{1} + y_3 X_1 + \omega_1 X_3 + \omega^*_1 X_4$$
$$M^2_L = \tilde{a}_6\mathbf{1} + y_6 X_1$$
$$M^2_E = \tilde{a}_7\mathbf{1} + y_7 X_1$$
$$A_E = \tilde{a}_8\mathbf{1} + \omega_5 X_2$$
$$A_U = \tilde{a}_4 X_5 + y_4 X_1 + \omega_2 X_6$$
$$A_D = \tilde{a}_5 X_1 + y_5 X_5 + \omega_3 X_2 + \omega_4 X_4$$

از آنجا که پارامترهای جرم شکست ابرتقارن هرمیتی هستند، در نتیجه $\tilde{a}_{1-3,6,7} > 0$ و مقادیر $x_1, x_2, y_1, y_3, y_6, y_7$ باید حقیقی باشند. بقیه ضرایب می‌توانند موهومی باشند. در اینجا $M_3$, $\tilde{M}_1$, $\tilde{M}_2$ پارامترهای مربوط به جملات جرمی گیجینو(gaugino) هستند، دو مورد اول جرم با توجه به فازهای نقض سی-پی یعنی $\phi_1$, $\phi_2$ موهومی هستند. $\mu$, $\phi_\mu$, $M_A$, $\tan\beta$ پارامترهای مربوط به قسمت هیگس-هیگسینو(Higgs/Higgsino)هستند.

با در نظر گرفتن مرتبه‌ی ۴ برای پارامتر طبقه‌بندی $O(\lambda^4) \sim O(10^{-3})$ تنها $x_{1-2}$ $y_1, y_3, y_6, y_7 \in \mathbb{R}$ و $\tilde{a}_{4,5,8}$, $y_{4-5} \in \mathbb{C}$ از معادله (۹) باقی می‌مانند، که ام-اس-اس-ام را با پارامترهای زیر می‌سازند:

(۹)
$$\tilde{M}_1 = e^{i\phi_1}M_1$$
$$\tilde{M}_2 = e^{i\phi_2}M_2$$
$$M_3, \tan\beta$$
$$\tilde{\mu} = e^{i\phi_\mu}\mu, M_A$$
$$M^2_U = \tilde{a}_2\mathbf{1} + x_2 X_1$$
$$M^2_Q = \tilde{a}_1\mathbf{1} + x_1 X_3 + y_1 X_1$$
$$M^2_D = \tilde{a}_3\mathbf{1} + y_3 X_1$$
$$M^2_L = \tilde{a}_6\mathbf{1} + y_6 X_1$$
$$M^2_E = \tilde{a}_7\mathbf{1} + y_7 X_1$$
$$A_E = \tilde{a}_8 X_1$$
$$A_U = \tilde{a}_4 X_5 + y_4 X_1$$
$$A_D = \tilde{a}_5 X_1 + y_5 X_5$$

---

[1] Supersymmetry breaking terms

در بررسی‌های جدید پارامترهای ام-اس-اس-ام۳۰ در این مقاله، سعی شده است که آنها را با توجه به نتیجه اخیر حد بالای ای-دی-ام الکترون، به روز رسانی شود. همه‌ی ۳۰ پارامترهای مدل مد نظر عبارتند از:

(۱۰)
$$\theta = \{M_1, M_2, M_3, Im(M_1), Im(M_2),$$
$$M_A, tan\beta, \mu, Im(\mu), \tilde{a}_{1,2,3,4,5,6,7,8}, x_{1,2},$$
$$y_{1,2,4,5,6,7}, Im(\tilde{a}_{4,5,8}), Im(y_{4,5})\}$$

برای کاوش در فضای پارامتر ام-اس-اس-ام۳۰، محدوده مقادیر برای هر یک از پارامترها در $\theta$ باید مشخص شود.

محدوده پارامترهای گیجینو ($M_1$، $M_2$) که مختلط هستند و $x_{1,2}$, $y_{1,3,6,7}$, در محدوده ۴- تا ۴ ترا الکترون ولت هستند و مقادیر $M_3$ از ۱۰۰ گیگا الکترون ولت تا ۴ ترا الکترون ولت می‌تواند باشد. پارامترهای $\tilde{a}_{1,2,3,6,7}$ ، $0 <$ درمحدوده ۱۰۰ گیگا الکترون ولت تا ۴ ترا الکترون ولت متغیر هستند. پارامترهای $\tilde{a}_{4,5,8}$, $y_{4,5}$, و $y_{4,5}$ در ۸- تا ۸ ترا الکترون ولت متغیر است. پارامترهای بخش هیگز این مدل توسط توده هیگز شبه‌نردبه‌ای $M_A$، که بین ۱۰۰ گیگا الکترون ولت تا ۴ ترا الکترون ولت می‌باشد، و پارامتر جفت دوگانه هیگز دوخطی $\mu$، در محدوده ۴- تا ۴ ترا الکترون ولت معتبر هستند. $tan\beta = \frac{<H_2>}{<H_1>}$ مجاز است بین ۲ و ۶۰ باشد. اینجا $<H_1>$ و $<H_2>$ مقادیر انتظار در خلاء برای هیگز دوبلت‌های ام-اس-اس-ام هستند.

برای هر کدام از پارامترهای ذکرشده، $P(\theta)$ ثابت[1] می‌باشد. از طریق اسکن کردن پارامترهای ام-اس-اس-ام۳۰ در فضای ۳۰ بعدی در بازه‌هایی که مشخص شده‌اند، تعدادی از نقاط در معادله (۵)، سازگاراست، جمع آوری شده است. ویژگی‌های مهم این نقاط نسبت به حد جدید گشتاور دوقطبی الکتریکی الکترون قابل دستیابی است، و این با استفاده از نتایج برازش کلی ام-اس-اس-ام۳۰ که در سال ۲۰۱۴ گزارش شده است و نمونه‌ی دیگری از نقاط ام-اس-اس-ام۳۰ اسکن شده‌ی تصادفی حاصل می‌شود [۱۱].

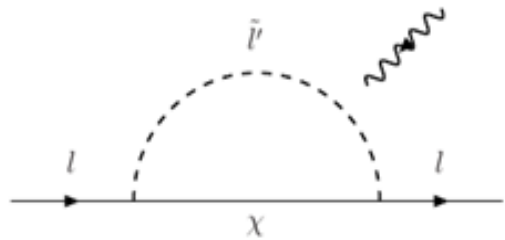

شکل ۶- نمودار فاینمن برای سهم X در ای-دی-ام لپتون l. طوری که سهم الکترون و نوترالینو و چارژینو[2] در نظر گرفته شده است.

شکل فوق نشان‌دهنده مشارکت چارژینو ام-اس-اس-ام و نترالینو در ای-دی-ام الکترون در نودار فاینمن می‌باشد [۲۰-۲۲] که توسط روابط زیر داده می‌شوند:

(۱۱)
$$\frac{d_e^{\chi^0}}{e} = -\frac{1}{16\pi^2}\sum_{im}\frac{m_{\chi_i^0}}{6m_{\tilde{e}_m}^2} Im[n_{im}^L n_{im}^R] F_2^N(x_{im})$$

$$\frac{d_e^{\chi^1}}{e} = -\frac{1}{16\pi^2}\sum_{k}\frac{m_{\chi_k^\pm}}{3m_{\tilde{\nu}_e}^2} Im[c_k^L c_k^R] F_2^C(x_k)$$

جایی که e بار الکتریکی الکترون است. در اینجا شاخص‌ها به ترتیب بر روی جرم نوترالینوها، اسلپتون‌ها و چارژینوها با مشخصه $i=1,2,3$ ، $m=1,2$ ، $k=1,2$ می‌روند. در این مقاله، برای حلقه‌ی تولید شده توسط نترالینو (چارژینو)، اسلپتون $\tilde{l}$ انتخابگر (یا اسنوترینو) برای مورد ای-دی-ام الکترون در نظر گرفته شده است. همچنین:

(۱۲)
$$x_{im} = m_{\chi_i^0}^2/m_{\tilde{l}_m}^2, x_k = m_{\chi_k^\pm}^2/m_{\tilde{\nu}_e}^2$$

$$n_{im}^R = \sqrt{2}g_1 N_{i1}\chi_{m2} + y_l N_{i3}\chi_{m1}$$

$$n_{im}^L = \frac{1}{\sqrt{2}}(g_2 N_{i2} + g_1 N_{i1})\chi_{m1}^* - y_l N_{i3}\chi_{m2}^*$$

$$c_k^R = y_l U_{k2}, c_k^L = -g_2 V_{k1}$$

با توجه به اینکه مقدار $y_l$ به صورت زیر می‌باشد:

(۱۳)
$$y_l = g_2 m_l/(\sqrt{2}m_W cos\beta)$$

به عنوان جفت‌شدگی یوکاوا برای لپتون l می‌باشند. ماتریس‌های اختلاط نوترالینو، چارژینو و اسلپتون به ترتیب N، U، V و X به گونه‌ای هستند که به صورت زیر می‌باشند:

(۱۴)
$$N^* M_{\chi^0} N^\dagger = diag(m_{\chi_1^0}, m_{\chi_2^0}, m_{\chi_3^0}, m_{\chi_4^0})$$

$$U^* M_{\chi^\pm} V^\dagger = diag(m_{\chi_1^\pm}, m_{\chi_2^\pm})$$

$$X M_{\tilde{l}}^2 X^\dagger = diag(m_{\tilde{l}_1}^2, m_{\tilde{l}_2}^2)$$

---

[1] Flat prior distributions  
[2] Chargino

و همچنین ضریب $F^{N,C}_{1,2}(x)$ با رابطه‌ی زیر بیان می‌شود:

(۱۵)

$$F^C_2(x) = -\frac{3}{2(1-x)^3}[3 - 4x + x^2 + 2\ln x]$$

$$F^N_2(x) = \frac{3}{(1-x)^3}[1 - x^2 + 2x \ln x]$$

فازهای سی-پی درگیر در بیان ای-دی-ام الکترون را می‌توان به شرح زیر استخراج کرد. ماتریس جرم نوترالینو با ماتریس زیر داده می‌شود:

(۱۶)

$$M_{\chi^0} = \begin{pmatrix} M_1 & 0 & -m_Z c_\beta s_W & m_Z s_\beta s_W \\ 0 & M_2 & m_Z c_\beta c_W & -m_Z s_\beta c_W \\ -m_Z c_\beta s_W & m_Z c_\beta c_W & 0 & -\mu \\ m_Z s_\beta s_W & -m_Z s_\beta c_W & -\mu & 0 \end{pmatrix}$$

طوری که متغییرهای ماتریس فوق به شکل زیر می‌باشند که در آنها $\theta_W$ زاویه اختلاط برهم‌کنش‌های ضعیف است.

(۱۷)

$$c_\beta = \cos\beta, s_W = \sin\theta_W$$

$$s_\beta = \sin\beta, c_W = \cos\theta_W$$

با استفاده از $D = diag\left(e^{-\frac{i\phi_\mu}{2}}, e^{-\frac{i\phi_\mu}{2}}, e^{\frac{i\phi_\mu}{2}}, e^{\frac{i\phi_\mu}{2}}\right)$ و معرفی ماتریس زیر:

(۱۸)

$$M'_{\chi^0} = \begin{pmatrix} M_1 e^{i\phi_\mu} & 0 & -m_Z c_\beta s_W & m_Z s_\beta s_W \\ 0 & M_2 e^{i\phi_\mu} & m_Z c_\beta c_W & -m_Z s_\beta c_W \\ -m_Z c_\beta s_W & m_Z c_\beta c_W & 0 & -|\mu| \\ m_Z s_\beta s_W & -m_Z s_\beta c_W & -|\mu| & 0 \end{pmatrix}$$

روابط زیر بدست می‌آیند:

(۱۹)

$$DM'_{\chi^0}D^T = M_{\chi^0}$$

$$\tilde{N}^* M'_{\chi^0} \tilde{N}^\dagger = diag\left(m_{\chi^0_1}, m_{\chi^0_2}, m_{\chi^0_3}, m_{\chi^0_4}\right)$$

جایی که $\tilde{N}^* = N^* D$ برقرار است.

---

[1] Stop-loop

به این ترتیب، ماتریس جرمی نوترالینو M' تنها دو فاز دارد که عبارت‌اند؛ $Arg(\mu M_1)$ و $Arg(\mu M_2)$، که $d_e(neutralino)$ به این‌ها بستگی دارد. به طور مشابه، $d_e(chargino)$ به فاز دوم سی-پی در بالا بستگی دارد.

سهم‌های ام-اس-اس-ام دیگری در ای-دی-ام وجود دارد، مانند مشارکت حلقه-توقف[1] در دو حلقه که نسبت مستقیم با $\tan\beta$ و نسبت معکوس با مربع جرم بوزون هیگز شبه نرده‌ای دارد [۲۳].

(۲۰)

$$d_e^{twoloop} = -\frac{e\kappa_e \alpha Y_t^2}{9\pi} \ln\left[\frac{M_{SUSY}^2}{m_A^2}\right] \sin(\theta_A + \theta_\mu)\tan\beta$$

که در آن $M_{SUSY}^2$ مقیاس جرم ابرتقارن است، $Y_t \sim 1$ جفت یوکاوا با کوارک بالا در مدل استاندارد است و $\kappa_e \sim 0.6 \times 10^{-25}$cm می‌باشد. بنابراین نیاز به جلوگیری از تولید بیش از حد ای-دی-ام الکترون، لزوماً به توزیع ام-اس-اس-ام۳۰ با مقادیر نسبتاً کمتر برای $\tan\beta$ و $m_A$ سنگین‌تر منجر می‌شود. این ویژگی را می‌توان در توزیع‌های پسین ام-اس-اس-ام۳۰ نشان داده شده در شکل ۷ مشاهده کرد.

در شکل(۷)، نمودار سمت راست، مقادیر گشتاور دوقطبی الکتریکی الکترون برای نقاط جمع شده ام-اس-اس-ام۳۰ را نشان داده می‌دهد با خط سیاه رنگ. کنار این، منحنی قرمز رنگ مربوط به نقاط ام-اس-اس-ام۳۰ که از برازش کلی در سال ۲۰۱۴ به دست آورده شد. حد جدید ای-دی-ام الکترون با خط عمودی نشان داده شد. قابل مشاهده است که اکثر نقاط اِسکن شده برای این مقاله با این حد سازگار است.

در نمودار وسط شکل(۷)، توزیع پسین پارامتر $\tan\beta$ آورده شده است. منحنی توزیع این پارامتر با توجه به قید جدید ای-دی-ام الکترون در مدل ام-اس-اس-ام۳۰ به رنگ سیاه نمایش داده شده است. می‌توان دید که این پارامتر در این حالت به سمت مقادیر نسبتا کمتر گرایش دارد. در نمودار سمت چپ هم توزیع جرم هیگز شبه اسکالر در مدل ام-اس-اس-ام۳۰ دیده می‌شود که قید ای-دی-ام جدید روی این جرم، مقادیر جرم را به مراتب بالاتر سوق می‌دهد.

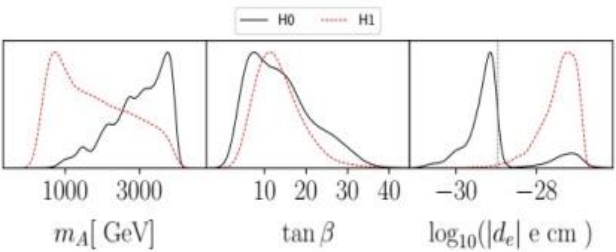

شکل ۷- توزیع یک بعدی برای جرم هیگز شبه اسکالر، $m_A$، (نمودار سمت چپ). نسبت مقادیر انتظاری خلاء برای هیگز دوتایی در مدل ام-اس-اس-ام که با $\tan\beta$ نشان می‌دهیم(نمودار وسط). و قید گشتاور دوقطبی الکتریکی، $d_e$،. (نمودار سمت

راست). منحنی $H_1$ ( قرمز) ، نتایج حاصل از برازش کلی ام-اس-اس-ام۳۰ در سال ۲۰۱۴ هست [۱۱] و منحنی $H_0$ (سیاه) نمایش همین مدل که با حد مجاز ای-دی-ام الکترون فعلی سازگار هست.

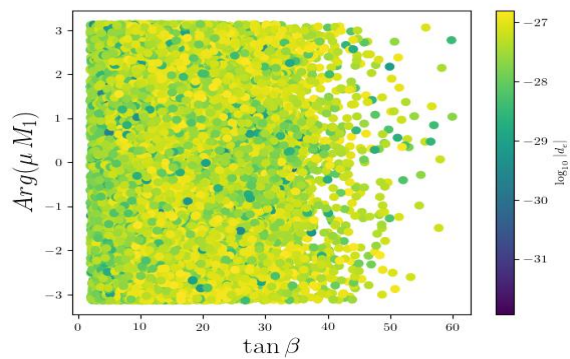

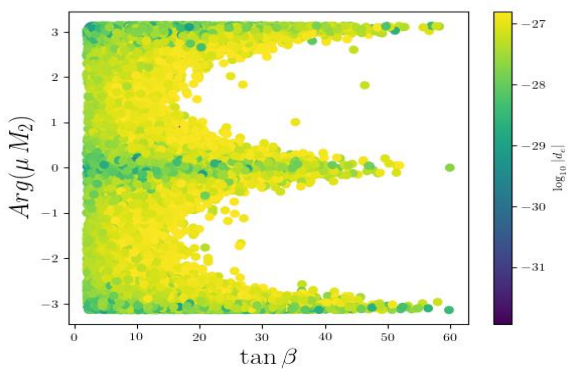

شکل ۸- همبستگی بین ای-دی-ام الکترون با فازهای سی-پی $Arg(\mu M_1)$، $Arg(\mu M_2)$ و $\tan\beta$ را نشان می‌دهد. این یک نمونه ام-اس-اس-ام۳۰ از یک تناسب جهانی با داده‌ها، از جمله چگالی بقایای ماده تاریک [۲۴] و ای-دی-ام الکترون [۲۶] می‌باشد.

شکل فوق که همبستگی ای-دی-ام الکترون، $|d_e|$، با فازهای سی-پی و $\tan\beta$ را با استفاده از تناسب جهانی نشان می‌دهد [۱۱] که این براساس نتایج همکاری آزمایش ACME-I است [۲۶].

مشاهده می‌شود که $|d_e|$ با $Arg(\mu M_2)$ همبستگی مستقیم دارد. نمونه نقاط با $|d_e| \sim 10^{-31}$ در اطراف $Arg(\mu M_2) = 0$ یا $Arg(\mu M_2) = \pi$ متمرکز شده‌اند، برخلاف مورد $Arg(\mu M_1)$ که مقادیر را در تمام محدوده $[0,\pi]$ شامل می‌شود حتی برای موارد کوچکتر از $|d_e|$ پیش‌بینی‌ها به این دلیل است که برای تناسب جهانی، چگالی بقایای ماده تاریک کاندید سبک‌ترین ماده‌ی تاریک ام-اس-اس-ام۳۰ باید با اندازه‌گیری پلانک چگالی بقایای ماده تاریک سازگار باشد[۲۴ و ۲۵] . محدودیتِ تراکم باقی‌مانده محتوای نوترالینو را مجبور می‌کند که عمدتاً دارای ترکیب هیگزینو[1] با حداقل محتوای بینو[2] باشد. به این ترتیب، فقط پارامترهای $\mu$ و $M_2$ در $M_\chi$ وارد

عبارت $d_e$ ارائه شده در بالا می‌شوند. از طرفی پارامتر $M_1$ توسط $|d_e|$ بدون محدودیت باقی‌می‌ماند. با اعمال کران بالای جدیدتر بر $|d_e|$ این ویژگی همان‌طور که در شکل ۹ نشان داده شده است باشد [۴ و ۱۳] ، باقی می‌ماند.

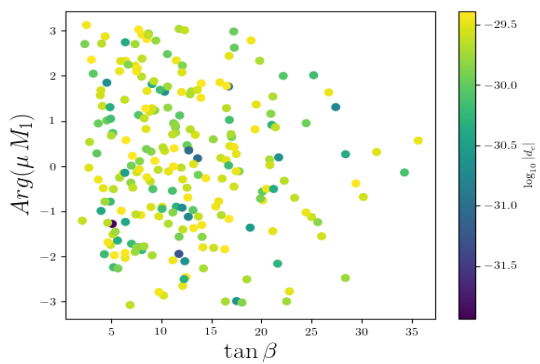

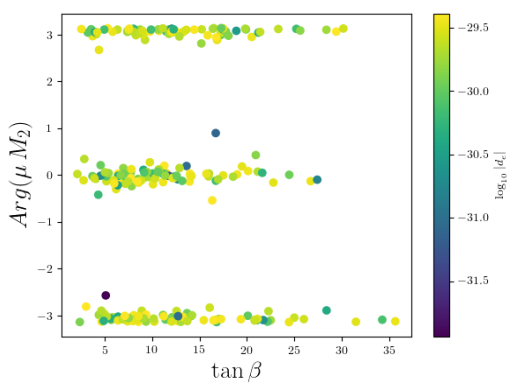

**شکل ۹-** همبستگی بین ای-دی-ام الکترون و فازهای سی-پی با $Arg(\mu M_1)$ و $Arg(\mu M_2)$ و $\tan\beta$. این از نمونه نزدیک ام-اس-اس-ام۳۰ از یک اسکن تصادفی با نقاط پذیرفته شده سازگار با اندازه‌گیری‌هایی از جمله چگالی باقی‌مانده ماده تاریک [۲۴ و ۲۵] و کران بالای ای-دی-ام الکترون به دست آمده است [۴ و ۱۳].

## ۵. نتیجه‌گیری

این مقاله، ما به صراحت نمونه‌ای از کار تحقیقاتی در زمینه پدیدارشناسی فیزیک ذرات ارائه کرده‌ایم که مستلزم مدل‌های نظری و محدود کردن آن‌ها با نتایج آزمایش‌ها است. حداقل مدل استاندارد ابر متقارن( ام-اس-اس-ام) و آخرین حد در گشتاور دوقطبی الکتریکی الکترون استفاده شد. اسکن‌هایی از مدل انجام داده‌ایم و مناطق فضای پارامتری را پیدا کرده‌ایم که با محدوده آزمایش سازگار است. اثر قید گشتاور دوقطبی الکتریکی الکترون را بر روی فضای پارامتری مدل ام-اس-

---
[1] Higgsino

[2] Bino

اس-ام ۳۰ این است که جرم شبه هیگز اسکالر را به مقادیر بالاتر نسبت به فضای پارامتر قبل سوق می‌دهد و $\tan\beta$ به مقادیر کمتر گرایش دارد.

تحلیل‌های ما بیشتر نشان می‌دهد که کران بالایی روی ای-دی-ام الکترون، فاز سی-پی آرگومان $Arg(\mu M_2)$ را به صفر یا $\pi$ فشرده می‌کند، برخلاف مورد $Arg(\mu M_1)$ که با وجود سازگاری با کران بالا، مقادیر را در تمام محدوده $[0, \pi]$ قرار می‌گیرد.

همبستگی قوی با سایر موارد قابل مشاهده مانند چگالی باقی‌مانده ماده تاریک با ال-اس-پی نوترالینو (LSP)[1] به عنوان نامزد ذره ماده تاریک وجود دارد. فیزیک موجودی که این را کنترل می‌کند، تأثیر سایر قابل‌مشاهده‌ها بر مناطق فضای پارامتری ام-اس-اس-ام را تحت تأثیر قرار می‌دهد. به عنوان مثال، برای جلوگیری از تولید بیش از حد ماده تاریک،

نترالینوها عمدتاً دارای وینو-هیگزینوس[2] هستند. این به نوبه‌ی خود بر نوع نترالینوها در حلقه‌هایی که به درون ای-دی-ام الکترون می‌روند، تأثیر می‌گذارد. طوری که این اثر به گونه‌ای است که فاز سی-پی $Arg(\mu M_1)$ به آرامی بدون محدودیت باقی می‌ماند. این نشان‌دهنده‌ی اهمیت تناسب جهانی است که به موجب آن اثر ترکیبی برخی مشاهدات و محدودیت‌های آزمایش‌های مختلف برای نگاشت مناطق سازگار در فضای پارامتری مدل با هم قرار می‌گیرند.

### تشکر و قدردانی

با تشکر فراوان از دکتر شانت شهبازیان و آقای عیسی محمدی هرگلان برای مطالعه نسخه قبلی و ارائه پیشنهادهایی در مورد ساختار و روند زبان مجله.

### مراجع

---

[2] Wino-Higgsinos

[1] Lightest supersymmetric particle; Neutralino